\documentclass[5p]{elsarticle} 

\usepackage{graphicx}
\usepackage{subfig}
\usepackage[version=3]{mhchem} 
\usepackage{dcolumn}
\newcolumntype{d}[1]{D{.}{\cdot}{#1} }
\usepackage{bm}
\usepackage{xcolor,soul}

\usepackage{siunitx}

\begin{document}
\title{Review: Solid-state physics of halide perovskites}
\date{\today} 

\author[jmf]{Jarvist Moore Frost}
\ead{jarvist.frost@imperial.ac.uk}

\address[jmf]{Department of Physics, King's College London, Strand, London WC2R 2LS, UK}
\address[jmf]{Department of Physics, Imperial College London, Exhibition Road, London SW7 2AZ, UK}

\begin{abstract}
Halide perovskite solar cells presented a unique opportunity to apply modern computational materials science techniques to an (initially) poorly understood new material.  In this review, we recount the key understanding developed during the last five years, through a narrative review of research progress.  The central enigma of the material is how it can be so defective, and yet work so well as a photovoltaic.  The physical properties of the material were understood through molecular and lattice dynamic calculations, revealing the material to show large dynamic responses on a wide range of time scales.  Longer length scales in the material was simulated with effective classical potentials, showing that complex domains can be generated by the interacting molecular dipoles, generating structured features in the electrostatic potential of the lattice.  Relativistic electronic structure reveals unique features in the bands, which may explain observed slow recombination, and could be used in high efficiency photovoltaics.  The large dielectric response of the lattice leads to a strong drive for the formation of polarons, some device physics of which are discussed.  These polarons offer a possible explanation for the observed slow cooling of photoexcitations in the material. 
\end{abstract}

\maketitle 

\section{Introduction}\label{introduction}

Halide perovskites burst into the consciousness of solid-state physicists with
the 2012 announcements\cite{Lee2012,Kim2012} of nearly \SI{10}{\percent} power conversion
efficiency solar cell devices. 
Mainstream silicon solar cells only recently passed \SI{25}{\percent}
efficiency, after 60 years of continuous development\cite{Green2009}.
That halide perovskite solar cells could reach such efficiencies with
relatively little research boded extremely well for the development of the
technology.  
The active material was solution processed methylammonium lead iodide, with no
vacuum or high-temperature steps in the preparation of the active material.
Along with the high crustal abundance of lead and iodine, the ease of lab-scale
synthesis indicated that low cost mass production would be possible. 

The choice of mesoporous electrical contacts was heavily influenced by the
germinal 2009\cite{Kojima2009} and 2011\cite{Im2011} use of the material as a
dye sensitiser with a liquid electrolyte in a solar cell. 
The working hypothesis was that light absorbed in this material would form
strongly bound excitons, charge transport would be slow, and recombination of
photoexcited states fast. 
One therefore needs to extract the charges as fast as possible, requiring
contacting on the nanometre scale. 

Yet the cells could also operate when on 
a non-conductive alumina scaffold\cite{Kim2012}. 
Planar devices\cite{Liu2013}, \SI{100}{\nano\metre} thick, had an estimated
\SI{1}{\micro\metre} photo-generated charge diffusion length\cite{Stranks2013}. 
Clearly photo-generated charges were capable of travelling a long way
through the material before recombining.

The enigma from a solid-state physics point of view was and is: how can
a solution-processed thin film be so efficient as a photovoltaic device? 
Solution-processed films are inevitably disordered and defective, yet a working
photovoltaic device requires a semiconductor with sufficiently high
charge-carrier mobility and low recombination velocity.

A new material with obvious technical importance presented an attractive
challenge to the materials modelling community. 
Little was initially understood about the material. 
Making the experiments and building consensus in what phenomena were being
observed would take time. 
By 2012 the materials modelling community had highly optimised codes and well
understood electronic structure methods based on the highly successful density
functional theory, and a wealth of computer time due to strategic investment in
high performance computing. 
We had a unique opportunity to try and understand the material from an
electronic structure perspective, and attempt to show the power of the
techniques by predicting the future of what was about to be measured, rather
than the less convincing argument of predicting past measurements on an
already well-studied material. 

\section{Hybrid solid-state materials}

Early electronic structure analysis of the material revealed unusual material
properties\cite{frost2014atomistic}. 
The static dielectric constant was large ($\epsilon=24.1$) as the material is ionic and soft.  
Intriguingly, the methylammonium itself has a permanent dipole of 2.29 Debye,
which appeared stable to the local polarisability\cite{frost2014atomistic}. 
This immediately implies, from knowledge that the resulting electrostatic
interaction between near neighbour methylammonium is of order of the thermal
energy, that a complex phase behaviour of dipole alignment will exist at room temperature. 
Ordering in these dipoles will result in macroscopic electric fields being
generated between phases of the material, which could lead to segregation of
the electron and hole charge carriers, and thus a reduction in the
recombination velocity. 
Band alignment of the material was similar to other photovoltaic
materials\cite{butler2015band}, indicating that a diversity of contact
materials was possible in future device architectures. 
Tuning of the organic hole collecting contact seemed to be one way of modifying
the open circuit voltage\cite{murray2015modular}. 

The simultaneous organic and inorganic nature of the material is a challenge
for modelling, not least because it requires researcher knowledge of both
molecular (vacuum) and solid-state (periodic) electronic structure methods.  
We have written a summary perspective on these challenges, suggesting
guidelines for accurate calculations\cite{whalley2017perspective}.


Solid-state electronic structure calculations usually rely on a periodicity of
the lattice. 
This enables the calculation of the macroscopic properties of a material, by
considering phase shifts within the Brillouin zone. 
However it relies on the perfect crystals (i.e. athermal) being
representative of the material under standard (thermal) conditions.

Given the unusual hybrid nature of the material, a first question to ask is:
what do the molecular dynamics look like? 
The first studies\cite{frost2014molecular} indicated a complex array of motion
on most timescales (a render of this data is available online\cite{frost2014mapiyoutube}). 
Though the material was a crystal, it was as soft as jelly. 
Further studies, as summarised in our 2016 review
article\cite{frost2016moving}, provided an essential answer that the soft
material exhibits thermally activated dynamic motion in almost every
conceivable manner. 

\begin{figure}
    \centering
    \includegraphics[width=0.85\columnwidth]{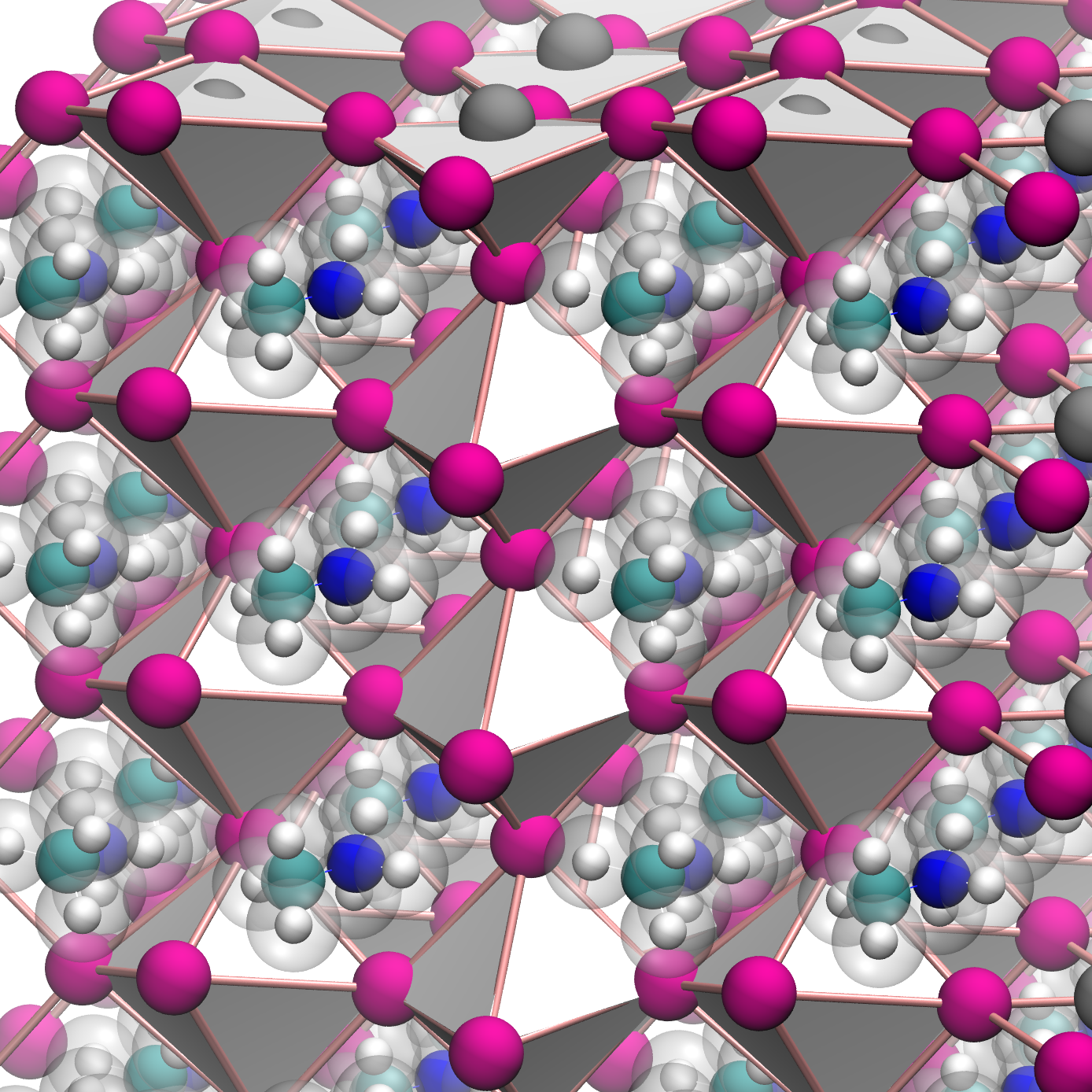}
\caption{Rendering of a finite-temperature distorted super-cell of methylammonium lead halide perovskite. 
Distortions around the relaxed position are generated with a frozen-phonon
approach, in this case evaluating an example of each of the first
6 branches of the phonon modes\cite{brivio2015lattice}, at the (001)
Brillouin zone location. Propagating these modes at their natural frequency
(the phonon eigenvalues) allows the relative motion and harmonics between the
different motion to be observed, without being obscured by the noise of an
actual molecular dynamics simulation.  Trajectories were generated with the
\textsc{Julia-Phonons} package, visualisation by \textsc{VMD} and the
\textsc{Tachyon} ray-tracer.}
\end{figure}

The materials are now believed to be mixed (ion and electron) conductors, and
that the interaction between this conduction leads to the unusual hysteresis in
device behaviour.  
Our early nudged elastic band calculations showed that vacancy mediated iodine
motion was likely to be the major contributor\cite{eames2015ionic}, providing
activation energies and associated predicted rates for multiple ion conduction
possibilities.

One interesting aspect of the molecular dynamics of perovskite structures is
that the supercell expansions used must be even. 
This is so that the well known perovskite tilting modes can be represented in the unit
cell. 
(Equivalently these give rise to soft phonon modes at the relevant high
symmetry point in the phonon Brillouin zone.)
All tin and lead halide perovskites appear to exhibit this
instability\cite{yang2017spontaneous}. 
Failing to include this point (for instance a threefold rather than double
expansion in one direction) leads to artificially constrained molecular
dynamics. 

By definition, molecular dynamics include anharmonic effects to all orders. 
However, the expansion of the super cell immediately quantises the phases of
periodic motion that can be represented. 

The main challenge of molecular dynamics is in interpreting the results, of
turning the trajectories of atomic location as a function of time into
something of scientific value. 
Qualitatively, the clear point was that the motion of the methylammonium
consisted of local wobbles, followed by discrete jumps in the orientation of
the methylammonium. 

Early density functional theory molecular dynamics extracted a timescale of the
motion\cite{frost2014molecular}, and by exploiting the $O_h$ symmetry of the
cubic unit cell, increased the signal to noise ratio of a measure of the
orientation of the methylammonium to show that the finite temperature ensemble
points at the faces (001) of the cube (along the diagonal of the octahedral
pocket), as could be inferred from the relative stability of this pose from
molecular statics calculations\cite{Brivio2013}. 
The contribution of the molecular dynamics is that it gives some idea of the
occupation by the thermal ensemble, and the distribution of parameters around
the minimum. 

From the molecular dynamics we can directly calculate the autocorrelation time
for the methylammonium motion. 
First analysis suggested a value of \cite{frost2014molecular,leguy2015dynamics}
of \SI{3.14}{\ps}.  
This compares to a value of \SI{14}{\ps} (with significant error bounds) from
quasi-elastic neutron scattering\cite{leguy2015dynamics}, and \SI{3}{\ps} from a
direct infrared measurement of individual methylammonium
ions\cite{bakulin2015real}.
Intriguingly, the timescale for reorientation of the larger formamidinium
(from molecular dynamics) molecule is actually \textit{faster} than
methylammonium at \SI{2}{\ps}, but with a much stronger preference for (001)
alignment\cite{weller2015cubic}. 
Mixed halide materials seem to result in locked octahedral structures (due to
the size mismatch), and immobilised organic cations\cite{selig2017organic}.


An alternative view of the dynamics of materials is to consider harmonic
expansions around the ground state structure, to calculate the normal modes of
the motion and a phonon band structure. 
A major challenge with hybrid lead-halide materials is finding this ground
state: the challenge of this procedure is in proportion to the condition number
of the dynamic matrix. 
Containing covalent hydrogens oscillating at \SI{90}{\THz} and soft lead-iodide
bonds with a natural frequency of \SI{1}{\THz}, this dynamic range makes
optimisation inefficient. 
Much effort is therefore required to locate the structural ground state, about
which the lattice dynamic perturbations can be constructed. 

Interpretation of these lattice modes\cite{brivio2015lattice} is assisted by
calculating the normal modes of the molecule in vacuum. 
This gives access to a classification, and infrared and Raman activity, that
can be compared to the mixed modes observed in the lattice dynamics simulation. 
This offers a mechanistic interpretation of the observed phonon band, in terms of the
well defined point-group theory and selection rules for a molecule. 

We followed this work by a joint experiment (Raman) and theory study, where we
attempted to pick out and identify every mode present in the dynamics, and
understand how the frequencies varied across the halide
series\cite{leguy2016dynamic}. 
Key results here were to identify the mid-energy torsional vibration of the
methylammonium C--N axis as being the molecular degree of freedom most affected
by the lattice constant (halide ion), understanding the mixed transverse-linear
optical modes generated by molecular coupling, and identify the `nodding
donkey' intermediate modes associated with the methylammonium rattling within
its cage. 
Finally the repercussions of the large frequency range in the polar optical
modes were considered.


%
%

\section{Monte-carlo models of large scale thermal fluctuations}

The methylammonium molecule possesses a permanent dipole of some
2.29 Debye. 
Electrostatic dipole-dipole interactions between these spins drive the
formation of a columnar anti-ferroelectric phase (spins head-tail aligned), as
revealed by a Metropolis Monte-Carlo lattice spin
simulation\cite{frost2014molecular}. 
This study also showed that the real-space extent of the fluctuations at room
temperature was on the order of 5 unit cells. 

These spin rotations do not occur in free space, but within a physical
lattice, the highly anisotropic $\mathrm{CH}_3-\mathrm{NH}_3$ molecules sitting
within a distorted octahedral pocket. 
In fact, the low temperature orthorhombic phase of the material is believed to
be one in which the octahedral pockets collapse into ordered perpendicular 
anti-ferroelectric terminations of the local spins. 
To a first approximation, the effect of this lattice confinement of the
methylammonium ions is a local strain term driving alignment of co-facial
methylammonium. 
This results in an effective dot-product energy term between nearest-neighbour
spins, which can be parametrised from the constrained rotation of an individual
methylammonium in a super cell as \SI{150}{\meV} which can be distributed
between the six co-facial neighbour spins\cite{leguy2015dynamics}. 
The resulting three-dimensional model is driven into a ferroelectric ground
state, with larger, deeper, basins of electrostatic potential. 
The model can also be extended to treat defects and mixed dipole
systems\cite{grancini2015role}. 

\begin{figure}
    \centering
    \includegraphics[width=0.95\columnwidth]{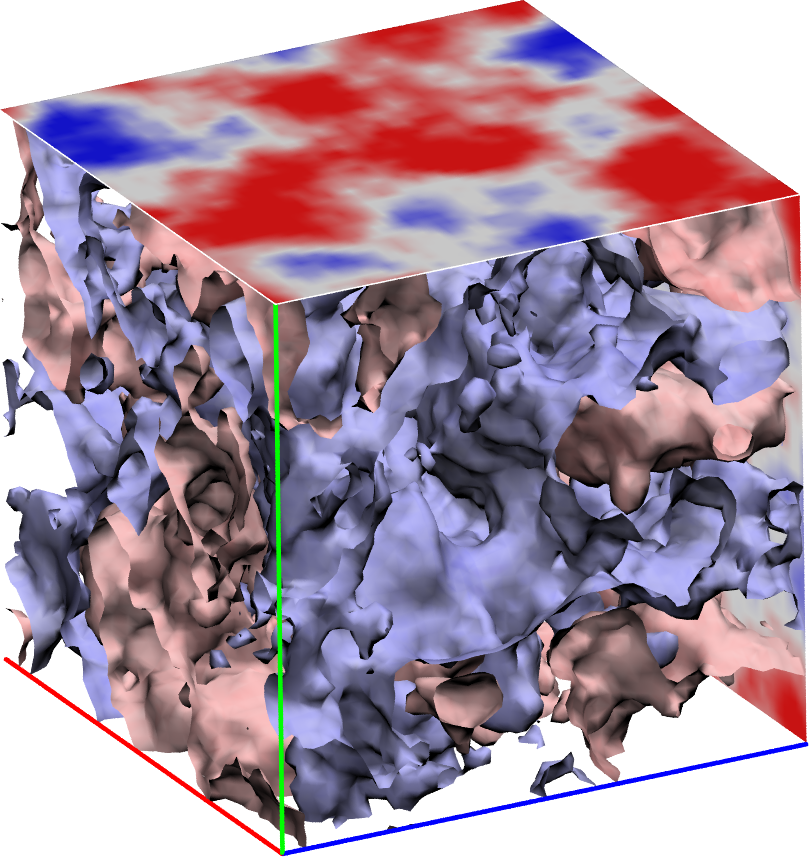}
\caption{Simulated electrostatic potential at 300 K in a high ferroelectric
driving force material, with a 50x50x50 supercell of material.  The
ferroelectric driving force gives rise to large regions of high (red) and low
(blue) electrostatic potential. The regions are continuous (percolating).
Simulated with the \textsc{Starrynight}
codes\cite{frost2014molecular,leguy2015dynamics}. }
\end{figure}

A challenge with such effective potential models is making strong predictions. 
The generated phase is the result of a careful balance of competing energy (enthalpy)
terms calculated from models and approximate electronic structure methods, to
which the entropic contribution is made directly by the Monte Carlo algorithm. 
Atomistic dynamics, even with empirical potentials\cite{Mattoni2015}, have difficulty in getting
to the necessary simulation length and time scales to investigate such domains. 
Thus it is perhaps better to think of these models as predicting possibilities,
with simulation temperature and the detailed values of the parameterisation to
be considered variables used to reproduce and compare to the experimental
reality. 

A more sophisticated parameterisation of the same functional form of model was later done by Tan et al.\cite{Tan2017}. 
More understanding of the local domain structure has been provided by Li et
al.\cite{Li2018} in terms of a `pair mode' analysis of the local cluster
structure. 

A shortcoming of these effective potential models is that they are wholly classical. 
Ma and Wang took a fully randomised methylammonium orientation supercell model
and provide clear indication via a linear-scaling fragment DFT method that
electrostatic potential fluctuations are present in the electronic structure\cite{Ma2014}.

Empirically, no room-temperature ferroelectricity is observed from bulk
electric field hysteresis measurements\cite{Fan2015}. 
Yet piezo-electric force microscopy reveal an applied-field structural
distortion which some authors have defined as evidence of
ferroelectricity\cite{Kutes2014}, columnar anti-ferroelasticity\cite{Rhm2017}
whereas others believe these indicate ferroelasticity only\cite{Hermes2016}. 
These domains appear to disappear upon heating, as would be expected from
melting the local dipole order\cite{Vorpahl2018}. 

There is currently no definitive evidence for whether there are nanometre
scale potential fluctuations in the perovskite material, which would most
directly affect the recombination kinetics of the photo generated charges.

\section{Device physics repercussions of large spin-orbit coupling}

Most materials modelling is with the density functional theory. 
A major weakness of this electronic structure method is that band-gaps are often under predicted. 
In the case of lead halide perovskites, the \SI{1}{\eV} band-gap under
prediction exhibited in a generalised-gradient approximation (GGA) functional is
almost exactly balanced by a large spin-orbit coupling that reduces the
band-gap by \SI{1}{\eV}. 
Forces are calculated from considering the occupied orbitals, so predicted
ground state structures, lattice dynamics and molecular dynamics should be
sufficiently accurate.
Relying on density functional theory for details of the unoccupied orbitals, as
is necessary for predictions of device behaviour, is more uncertain. 

Diffraction measurements of the high temperature phases of inorganic halide
perovskites solve to a cubic structure. 
Due to the inversion symmetry in such a structure, the Rashba effect cannot be
supported and ferroelectricity cannot be present. 
However these coherent measurements are insensitive to disorder unless it has
long range correlation. 

An early high quality electronic structure calculation\cite{Brivio2014} used
a full treatment of spin-orbit coupling along with a self-consistent
partially-screened exchange interaction (QS\textit{GW}). 
The unit cell was a relaxed `pseudo-cubic` ferroelectric construction, where
finite-temperature disorder was approximated by allowing the cell to distort
and the ions to relax. 
This revealed an intriguing feature in the conduction band---a 
Rashba-Dresselhaus splitting had occurred with the spin polarised electrons
being projected (in opposite, antipodal directions) away from the high-symmetry
extrema. 

This has obvious and severe consequences for the solar-cell device
characteristics. 
The band-gap becomes very slightly indirect---at low fluences the photo-excited
electrons will relax into these Rashba pockets, where due to the offset in
reciprocal space, there is no population of holes to recombine directly with. 
Effectively you have paid with enthalpy to reduce the recombination rate. 
A simple model based on thermal excitation back to the high-symmetry location
and then direct emission (ignoring indirect transitions), predicts a 350 fold
reduction in recombination due to this effect\cite{Azarhoosh2016}, with strong
temperature and fluence dependent behaviour. 
A device level analysis by Kirchartz and Rau\cite{Kirchartz2017} suggests that
a material with an indirect and direct gap would have an increased minority
carrier lifetime, but not an increased open circuit voltage. 

A second effect, as pointed out by Zheng et al.\cite{Zheng2015} is that the
spin-symmetry of the electron and hole states may be incorrect for direct
recombination.  
Combined, these two factors present in the bulk reciprocal space electronic
structure may be part of the reason why recombination of minority carriers is
is so slow in these materials. 
They both arise as a result of large spin orbit coupling, interacting with a 
locally broken symmetry environment. 
Such symmetry breaking has been been observed by inelastic X-ray
scattering\cite{beecher2016direct,Comin2016} for the methylammonium lead halide
material. 

By combining molecular dynamics and frozen-phonon structures from lattice
dynamics, with an electronic structure analysis of the Rashba splitting in the
resulting structures\cite{mckechnie2018dynamic}, we found that the nature of
spin-splitting in both the methylammonium and cesium lead-halide perovskites
was qualitative identical at room temperature. 
Thermal fluctuations in  all lead-halide perovskite structure are
sufficient to generate these states, as the material is soft leading to
distortions of the octahedra generating local polar fields intersecting with
the frontier orbitals. 

Measurements to confirming or contradict these suggestions from theory have
been relatively slow. 
The total Brillouin-zone displacement is relatively small, and the materials
themselves are not very stable under vacuum.  
The Rashba-Dresselhaus effect is driven by a local field. 
ARPES measurements on cleaved samples often sample surface effects; 
giant splitting was observed in the valence band\cite{Niesner2016}, the
presence of which in the bulk is not supported by any electronic structure
studies.

A study of the circular galvanic effect in methylammonium lead iodide
samples suggests that a spin-split indirect-gap exists at room temperature with
a stabilisation energy of 110 meV\cite{Niesner2018} below the direct gap. 
The positive temperature dependence of the magnitude of the spin-split
indicates that it may be generated by the dynamic fluctuations discussed above. 
The experimental value compares to predictions of 75 meV\cite{Azarhoosh2016} for the athermal
pseudo-cubic methylammonium unit cell, and 20 meV\cite{mckechnie2018dynamic}
for thermal ensembles of both the cesium and methylammonium. 

\section{Feynman variational polaron physics}\label{polaron}

Polarons as quasi-particles consisting of a charge carrier dressed in polar phonon excitations. 
, the coupling of a charge carrier to a polar phonon mode, 
results in electron confinement.  
Such localisation provides a natural link between a delocalised band structure
(where every electron is partially in every unit cell), and a more mechanistic
device-physics picture of individual electrons discretely moving through
a material. 
The driving force for polaron formation is the mismatch between the optical and
low-frequency dielectric constants; the ionic contribution from polar phonon
modes. 
The polarisation response of the lattice beyond the charge-carrier wavefunction
envelope generates a dip in the electrostatic potential which attempts to
localise the charge-carrier. 
The localisation renormalises the effective mass, and changes the nature of charge                                           
transport, recombination, and response to inhomogeneities in the material.

As the halide perovskites are ionic and soft, this driving force is large, 
$\epsilon_{opt}=4.5$ vs $\epsilon_{stat}=24.1$\cite{frost2014atomistic}. 
Immediately this implies that polarons are likely a good representation of the
charge-carriers in the material, rather than plane-waves. 

The Fr\"ohlich polaron is a simple model Hamiltonian with an effective-mass
electron coupled to a polar phonon mode. 
The dimensionless Fr\"ohlich parameter $\alpha$ quantifies the 
dielectric electron-phonon coupling, 

\begin{equation}
\alpha =
    \frac{1}{4\pi\epsilon_0}
    \frac{1}{2}
    \left ( \frac{1}{\epsilon_{opt}}-\frac{1}{\epsilon_{stat}} \right )
\frac{e^2}{\hbar\Omega}
\left( \frac{2m_b \Omega}{\hbar} \right)^{1/2}
.
\end{equation}

Here the variables $\epsilon_{opt}$ and $\epsilon_{stat}$ are the optical (high frequency)
and static dielectric constants (expressed in units of the permittivity of
vacuum, $\epsilon_0$), $\Omega$ is the angular frequency of the polar phonon
mode, $m_b$ is the bare band effective mass. 
The other physical constants are $\hbar$ the reduced Planck constant, $e$ the
electron charge. 

In 2014, with a knowledge of the QS\textit{GW} effective mass\cite{Brivio2014} $m_e=0.12$ (in
units of the bare electron mass), the aforementioned dielectric constants, and
a guess at the typical semiconductor polar phonon frequency (\SI{9}{\THz}) this
gives rise to a Fr\"ohlich $\alpha$ coupling parameter of
1.2\cite{frost2014molecular}. 
The most successful solution of the Fr\"ohlich Hamiltonian was the Feynman variational approach\cite{Feynman1955}. 
Here the Coulombic interaction of the charge-carrier with its lattice
polarisation field, is remapped to an exactly solvable (by path integration)
model of a harmonic interaction (characterised by the variable $v$), decaying
in time (exponential decay set by the variable $w$). 
These $v$ and $w$ parameters (which are expressed in units of the angular
frequency of the phonon mode $\Omega$) can be freely varied to minimise the mismatch in
this trial action, and the full Coloumb action, variationally approaching the true state. 

For such small values of $\alpha$ (mainly due to the light effective mass successfully resisting
confinement), an asymptotic estimate of the variational solution ($w$ fixed at
3, $v$ linear increased with $\alpha$) and a  perturbative treatment of the
polaron energy (and so, effective mass) is sufficient. 
This predicts an athermal effective mass renormalisation of
$+25\%$, and a polaron radius of 5 unit cells\cite{frost2014molecular}. 
Considerable extra effort by Schilpf et al.\cite{Schlipf2018}, integrating the
individual phonon contributions across the Brillouin zone, but using the same
asymptotic perturbative limit of the Feynman polaron theory, improves this
estimate to $\alpha\,=\,1.4$ and effective mass renormalisation of $+28\%$. 

The estimates of polaron extent (see equation 2.4 in \cite{Schultz1959}) come
from the width of the Gaussian wavefunction which is a ground state to the
harmonically bound electron and fictitious mass. 
This is of key importance when considering the influence of inhomogeneities
in the real-space lattice. 
The charge-carrier state samples the lattice over the extent of its
wavefunction. 
Fluctuations on a scale much smaller than the extent of the charge-carrier are integrated out. 

\section{Polaron mobility}

To go beyond these perturbative athermal effective-mass renormalisations and
directly predict the polaron features and dynamics (most importantly the
temperature-dependent charge-carrier mobility), we needed to return to the
original theories of Feynman et al.\cite{Feynman1962} and write computer codes
to fully solve the equations of motion\cite{MooreFrost2018}. 
These codes undertake the whole computational pipeline for Feynman variation
polarons for an arbitrary material, fully optimising the variational parameters
by auto-differentiation of the (temperature dependent) free energies, and
contour integrating the polaron response function to predict finite-temperature
polaron mobility and other observables. 

Reliable lattice dynamic calculations were slowed by the
difficult of establishing a reliable ground-state structure about which to
perturb the structure. 
Once this work was done, the infrared activities of individual modes can be
arrived at by considering propagation of the Born effective charges along the
gamma-point phonon eigenvectors\cite{brivio2015lattice}. 
The wide spectrum of polar modes with frequencies as low as \SI{1}{\THz}
($\hbar\omega\,=\,$\SI{4}{\meV}) immediate suggests a complex
temperature-activated behaviour as the polar phonon scattering modes
individually start scattering as a function of charge-carrier energy and
therefore temperature (see Fig. 10 in Ref. \cite{leguy2016dynamic}). 
How should you marry this spectrum with the Fr\"ohlich Hamiltonian single
frequency mode? 

The ultimate answer may be to extend the Feynman variational method, and
explicitly consider the simultaneous action of these multiple modes
numerically in the calculation of the action. 
Hellwarth et al.\cite{Hellwarth1999} provide two averaging techniques
to arrive at effective frequency mode, a thermal `A' technique, and an easier
to apply athermal `B' technique. 
This `B' technique gives an effective optical mode frequency of
\SI{2.25}{\THz}\cite{frost2017calculating}. 
Combined with the effective-mass and dielectric constants as above, this gives
rise to an Fr\"ohlich $\alpha$ of 2.4 (for electrons in methylammonium lead
iodide, see Ref. \cite{frost2017calculating} for further perovskite material
parameters).  

Such values of $\alpha$ are in the intermediate regime between weakly coupled
(perturbative) and strongly coupled (localised, hopping) polarons. 
Interestingly, by virtue of their stiffer phonons and lighter effective masses,
tin based halide perovskites have much smaller alpha parameters, and therefore
should have less localised (more band like) charge
carriers\cite{frost2017calculating}.

Therefore we need a temperature-dependent and non-perturbative mobility
measure. 
The original Feynman path-integral mobility
theory\cite{Feynman1962} offers exactly this, at the cost of numerically
contour integrating the polaron linear-response function. 

Applied to halide perovskites, this gives parameter-free
temperature-dependent mobilities\cite{frost2017calculating} in good agreement with
(room-temperature) single-crystal measurements. 

Though these predicted mobilities are numerically calculated, they are
relatively featureless and monotonic, due to the athermal effective polar mode
frequency. 
Again, a more complete result would be to extend the theories to enumerate the
individual scattering modes. 

One limitation of this polaron model is that the effective phonon mode is
harmonic (and therefore has an infinite lifetime). 
In the polaron theory this leads to a resonance of energy between the electron
and phonon degrees of freedom. 
Slower polar reorientation degrees of freedom, as would be expected from the
methylammonium rearrangement, would not be harmonic. 
Generally the phonon modes in hybrid lead halides are predicted to be highly
anharmonic\cite{whalley2016phonon}, for which there is now some direct
experimental evidence\cite{gold2018acoustic}. 
This anharmonicity broadens the spectral response, and decreases the infinite 
phonon lifetime to a sub-picosecond timescale. 
Rather than a resonant store of energy in the polaron model, these are likely
to be an additional dispersive loss mechanism. 
The polaron theories have not yet been extended to deal with anharmonicity. 

Other methods of predicting mobilities in these materials build on theories
(small polaron hopping; free-electron scattering) which are not valid for the
intermediate behaviour of most solid-state polar materials. 
They often rely on empirical parameters (such as a scattering time), which
limits predictive power. 
Combined these problems lead to poor absolute accuracy, and incorrect
predictions of temperature-dependence.

The temperature dependence of mobility ($\mu \propto T^n$) can give
circumstantial evidence for the nature of the scattering process controlling
(limiting) mobility (see chapter 3 in Ref. \cite{Ridley}).  
Emphasising that such analysis is approximate and based on semi-classical
models of mobility, acoustic deformation-potential and non-polar optical phonon mode 
give $\mu \propto T^{-\frac{3}{2}}$, whereas polar optical phonon mode scattering
and acoustic phonon piezoelectric scattering give $\mu \propto T^{-\frac{1}{2}}$. 
Also note actually getting from the raw measurement to a mobility measure relies on
a model of conductivity, most often the classical gas Drude model.
Practically speaking the values are derived by fitting a straight line to the
mobility displayed on a log-log axis.
 
Hall-effect measurements in methylammonium lead bromide single crystals
indicate an exponent of -0.5 in the low temperature phase, and -1.5 in the high
temperature phase. 
Time-resolved terahertz conductivity measurements on
methylammonium lead halide fit a $\mu \propto T^{-\frac{3}{2}}$
relationship\cite{Milot2015}, but discounting the low temperature orthorhombic
phase data and fitting to a power law in the linear (non logarithmic) metric
gives a trend of -0.95\cite{frost2017calculating}. 
This is in much closer agreement to the $-0.5$ from the numeric polaron
theory\cite{frost2017calculating}.  
Measurements of temperature dependent mobility on formamidinium lead iodide
perovskite\cite{DaviesFAPI2018} reveal a $-0.53$ exponent, in almost direct
agreement with the polaron model, and semi-empirical polar mode scattering.

Where does the extra scattering in the methylammonium material come from? 
One possibility is that is due to the formation of polar domains and resulting fluctuations in
electrostatic potential\cite{frost2014molecular}.
Alternatively it may be the direct dielectric response of this
additional dispersive polarisation response, discussed as a dielectric
drag\cite{Bonn2017}. 
Both of these explanations invoke a unique role of the permanent 2.29 Debye
dipole in methylammonium\cite{frost2014atomistic}, but with a different
mechanistic effect---either the dynamic interactions leading to a mostly-static
potential landscape, or the dynamics direct interacting with the charge
carriers. 
The two models could be distinguished by their different temperature response. 
It is not clear why the `dielectric drag' would be greater at higher
temperature, as required to increase the methylammonium temperature dependent
exponent, when the dielectric response of a polar liquid \textit{decreases} as
$\frac{1}{T}$ with temperature\cite{Kirkwood1939} (though the rate of response
would increase with temperature).
Further comparative mobility measurements of methylammonium, formamidinium and
inorganic cations could inform on whether the polar-methylammonium hypothesis holds.

Overall it seems likely that a deeper understanding of transport in these
materials will require a closer interaction between experiment and theory.
Rather than approaching the same final phenomenological quantity of mobility
from opposite directions, modelling of experimental observables would
simultaneously validate the model selection, and enable stronger statistical
statements to be made from both sides.

\section{Slow cooling of photoexcited states}

Another unusual behaviour noticed in the hybrid perovskite materials is slow
cooling (thermalisation times exceeding nanoseconds) upon above band-gap
photoexcitation\cite{Price2015}. 
This is of technical interest as such slow cooling is a prerequisite for being
able to extract charges fast enough to create a hot-carrier solar cell. 
The measurements are challenging to interpret as there are multiple bands than
can be excited in the halide perovskites, and often the bright laser light used
to get signal in the measurement takes the material into a photogenerated
charge density vastly larger than that reached by a solar cell device in the
dim sun, even under concentration. 

The polaron model, along with the understanding that the hybrid halide
perovskites have exceptionally low thermal mobility due to anharmonic coupling
between the phonon modes\cite{whalley2016phonon}, provides a simple mechanistic
explanation\cite{frost2017slow}. 
The polaron localisation leads to a localised heating, due to the low lattice
thermal conductivity, this energy takes a long time to dissipate. 
Thermal conductivities being ten times larger in the fully inorganic materials
precludes this behaviour. 
The threshold `phonon bottleneck' transition at higher fluence into a faster
cooling regime can be understood in terms of the density at which the localised
polarons start to overlap and lose their character. 
The material-specific localisation of the polaron gives rise to a quantised
behaviour (i.e. some characteristic thermalisation time dependent only on the
material, and an initial polaron temperature dependent on band-gap exceeding
energy\cite{Hopper2018}). 

Recently a measurements by a technique where the (thermalised) photo-excited states are
reheated by a delayed infrared pulse allowed careful observation across
a systematic range across the halide perovskite family\cite{Hopper2018}. 
Here the variations in cooling rate and density-dependence are explained by a
simple mechanistic model considering the specific heat capacity (and therefore
temperature reached in the localised polarons), and the characteristic
timescale of optical phonon scattering (from the polaron theories) assumed
proportionate to a cooling rate dominated by emission of optical phonons.

\section{Higher lying excited states}

One intriguing feature of the lead halide perovskites are the higher lying
excited states, attainable with violet light (3.2 eV) in the case of the lead
iodide perovskite. 
These states are mainly composed of Pb-6p orbitals, the same ones that make up the conduction band. 
But due to the spin orbital interaction, these orbitals have been pushed upwards in energy. 
Combined with the Rashba spin-splitting leading to incommensurate locations in
reciprocal space, this suggests that an intermediate band photo
ratchet\cite{PhotonRatchet2012} may be present as a bulk electronic structure
effect in the material\cite{frost2016photon}.  

Such a mechanism would not be directly useful to make Shockley-Queisser limit
exceeding photovoltaics, as the gaps in lead-halide perovskites are too high to
usefully divide up the solar spectrum.  

The lead-tin 50:50 alloy perovskite exhibits a much reduced 1.2 eV gap, 
and may offer a genuine route to higher efficiency. 
But the indirect band structure (therefore low oscillator strengths, therefore
low photoexcited charge densities) may preclude the effect from being
active\cite{Goyal2018}.  
Similar features may be present in other heavy-atom semiconductors with broken
inversion symmetry, leading to a potential new class of relativistic
semiconductors\cite{Butler2016,butler2015ferroelectric}. 

Such higher lying states are optically accessible, and have been characterised
by both ellipsometry measurements, and electronic
structure\cite{Leguy2016ellipso}. 

Recently a fluence-independent quantum-yield of exactly 2 has been observed in CsPbI3
nano-crystals upon 4.0 eV photo excitation\cite{deWeerd2018}. 
Though this is attributed by the researchers to multiple carrier generation,
the strikingly quantised quantum efficiency suggests that it may instead be sequential
emission on the way down the photon ratchet. 
Very strong Auger recombination in these materials has been attributed to
excitation to this higher energy conduction band\cite{Shen2018}, which may also
be a route for carrier multiplication at high photoexcited densities. 

Interesting photophysics are suggested by the relativistic band structure;
interesting photophysics are being observed in the laboratory. 
Whether the theory explains the observations, and whether these behaviours can
be made use of in a practical device, is not yet clear. 

\section{Conclusion}

The five years of computational halide perovskite described in this review have
benefited enormously from deep collaborations with a wide range of other
research groups, both computational and experimental. 
Hypotheses and new ideas can be quickly and efficiently developed in pure
computational studies, but the true understanding in this field has come from
experiment and theory working together to understand the observed data. 

We have not discussed any of the practical issues of developing
a working photovoltaic technology. 
The two chief challenges in this area have been the stability of the material,
and the fundamental economic case for switching process, to still have a
standard Shockley-Queisser single band-gap technology. 
Much practical work now goes into tandem-cell applications, where reduced
stability compared to a standard technology is a price worth to pay for the
extra efficiency. 
 
The long term benefits of the halide perovskite electronic structure studies may
come from the fundamental insights offered into the observed low
minority-carrier recombination rate. 
The two main suggestions are that this is either due to the relativistic
electronic structure leading to spin-split Rashba pockets (or spin selection
blocked recombination), or that the soft polar nature of the material leads to
recombination pathways reduced by real-space segregation, dielectric screening
and polaron effects. 
Further development and understanding of these processes may give guidance as
to what other semiconductors can be concocted from the periodic table to give
similar beneficial behaviour, without the relative instability of the halide
perovskite materials. 

\textbf{Acknowledgement}
We thank Piers Barnes, Jenny Nelson, Mark van Schilfgaarde and Aron Walsh for many years of stimulating discussions. 
Much of this work was undertaken on EPSRC Grant EP/K016288/1. 
J.M.F. is supported by EP/R005230/1. 


\bibliographystyle{elsarticle-num}
\bibliography{RisingStar,PublicationsFull}

\end{document}